\documentclass{JHEP3}
\preprint{LU-TP 03-47\\
  hep-ph/0311252}

\usepackage{cite}
\usepackage{epsfig}
\usepackage{color}

\usepackage{graphics}
\usepackage{axodraw}
\usepackage{inputenc}
\usepackage{xspace}
\inputencoding{latin1}
 \renewcommand\email[1]{{\scriptsize\tt\href{mailto:#1}{#1}}}
\newcommand{\gtaet}{\raisebox{-0.8mm}%
{\hspace{1mm}$\stackrel{>}{\sim}$\hspace{1mm}}}

\newcommand{\abar}{\ensuremath{\overline{\alpha}}}

\newcommand{\kT}{\ensuremath{k_{\perp}}}

\newcommand{\cascade}{C\scalebox{0.8}{ASCADE}\xspace}
\newcommand{\ldcmc}{\scalebox{0.8}{LDCMC}\xspace}
\newcommand{\leading}{\textit{leading}\xspace}
\newcommand{\gluonic}{\textit{gluonic}\xspace}
\newcommand{\standard}{\textit{standard}\xspace}
\def\mrm#1{\mathrm{#1}}
\def\sub#1{\ensuremath{_{\mrm{#1}}}}
\def\sup#1{\ensuremath{^{\mrm{#1}}}}

\newcommand{\eqref}[1]{eq.~(\ref{#1})\xspace}
\newcommand{\eqsref}[1]{eqs.~(\ref{#1})\xspace}

\skip\footins = 1\bigskipamount plus 2pt minus 4pt                              

\title{\boldmath Central Exclusive Scalar Luminosities\\ from the
  Linked Dipole Chain Model Gluon Densities }

\author{Leif Lönnblad and Malin Sjödahl\\
  Dept.~of Theoretical Physics,
  S\"olvegatan 14A, S-223 62  Lund, Sweden\\
  E-mail: \email{Leif.Lonnblad@thep.lu.se}
    and \email{Malin.Sjodahl@thep.lu.se}}
  
  \abstract{We investigate the implication of uncertainties in the
    unintegrated gluon distribution for the predictions for central
    exclusive production of scalars at hadron colliders. We use
    parameterizations of the \kT-unintegrated gluon density obtained
    from the Linked Dipole Chain model, applying different options for
    the treatment of non-leading terms. We find that the luminosity
    function for central exclusive production is very sensitive to
    details of the transverse momentum distribution of the gluon
    which, contrary to the \kT-integrated distribution, is not very
    well constrained experimentally. }

\begin{document}
 
\sloppy
 
\section{Introduction}
\label{sec:intro}

To detect the Higgs at hadron colliders such as the Tevatron or the
LHC is far from a trivial task. Especially if it is rather light and
predominantly decays into bottom quarks, the background from standard
QCD processes is huge, making the expression ``needle in a haystack''
seem like a severe understatement. Looking for Higgs signals in the
relatively clean environment of diffractive events is therefore an
appealing prospect, provided the cross sections are sufficiently high.
Several suggestions for what kind of diffractive processes could be
used and how to calculate the corresponding cross section for the
Higgs and the background have been made
\cite{Cudell:1996ki,Levin:1999qu,Khoze:2001xm,Cox:2001uq,
  Boonekamp:2002vg,Enberg:2002id}.

The cleanest and most promising process is usually referred to as
central exclusive Higgs production, pp$\rightarrow$p$+$H$+$p (where
the $+$ symbolizes a large rapidity gap), and was suggested by Khoze,
Martin and Ryskin (KhMR)\footnote{We shall here refer to their
  calculation as KhMR to distinguish it from the KMR procedure for
  obtaining unintegrated gluon densities from integrated ones by
  Kimber, Martin and Ryskin\cite{Kimber:2001sc}.}\cite{Khoze:2001xm}.
This process has several advantages. If the protons are scattered at
small angles with small energy loss and they are detected in very
forward taggers, the centrally produced system is constrained to be in
a scalar state, which reduces the background from e.g.\ normal QCD
production of b-jets. By matching the mass of the central system as
measured with the central detectors, with the mass calculated from the
energy loss of the scattered protons, it is possible to exclude events
with extra radiation outside the reach of the detectors.

To calculate the cross section for this process one starts off with
the standard gg$\rightarrow$H cross section and adds the exchange of
an extra gluon to ensure that no net colour is emitted by the protons.
Then one must make sure that there is no additional radiation what so
ever in the event, which gives rise to so-called soft and hard
gap-survival probabilities. The soft survival probability ensures that
the protons do not undergo any additional soft rescatterings, while
the hard survival probability ensures that the there is no additional
radiation from the exchanged gluons. Since the probability to emit
really soft gluons diverges, it is necessary to introduce some natural
cutoff, in order for the latter survival probability to remain finite.
This is accomplished by letting the exchanged gluons have finite
transverse momenta so that soft gluons cannot resolve the individual
colour flows in the total colour singlet exchange. These transverse
momenta must compensate each other so that the net transverse momenta
of the scattered protons are zero. This means that the two gluons
emitted from each proton are highly correlated and it is necessary to
introduce so-called off-diagonal, or skewed, parton densities
(odPDFs), which in addition must be \kT-unintegrated
(oduPDFs\footnote{Throughout this paper we shall use the following
  abbreviations: PDF refers to the standard diagonal integrated
  parton (gluon) density, uPDF is the diagonal \kT-unintegrated
  density, odPDF is the off-diagonal integrated and oduPDF is the
  off-diagonal \kT-unintegrated density.}). With this formalism it is
then possible to factorize the central exclusive production of any
scalar resonance, $R$, into the standard partonic gg$\rightarrow R$
cross section multiplied by a gluon luminosity function which includes
both the additional gluon exchange and the gap-survival
probabilities.  In this way we can turn any hadron collider with
forward taggers into a kind of colour-singlet gluon collider with
variable center of mass energy.

There are several uncertainties associated with this process. Both
theoretical ones, such as how to calculate the soft survival
probability, and experimental ones, such as how well the scattered
protons can be measured. In this paper we will concentrate on another
theoretical uncertainty, namely how well we know the oduPDFs which
enters to the fourth power in the cross section. The quoted PDF
uncertainty in \cite{Khoze:2001xm} is a factor of two\footnote{In
  later papers the quoted uncertainty is factor $2.5$ up or down
  \cite{Kaidalov:2003fw,Kaidalov:2003ys}}, which may seem large, but
we will here argue that the uncertainty may be even larger.

The factor of two uncertainty was obtained by using a particular
procedure to obtain the gluon oduPDF from the standard diagonal
integrated gluon PDF, and then using different parameterizations for
the latter. The problem with this estimate is that the diagonal
integrated gluon PDF is fairly well constrained experimentally, while
the diagonal unintegrated one is not, and the off-diagonal
unintegrated even less so. In this paper we will use an alternative
way to obtain the gluon uPDF, based on the so-called Linked Dipole
Chain model \cite{Andersson:1996ju,Andersson:1998bx}, which is a
reformulation of the CCFM \cite{Ciafaloni:1988ur,Catani:1990yc}
evolution for uPDFs. With the LDC model the uPDFs can in principle be
better constrained since it is possible to compare with less inclusive
experimental data, looking at details of the hadronic final states of
events. Especially observables such as forward jet rates in DIS should
be sensitive to the actual \kT-distribution of gluons in the proton.
Unfortunately it turns out to be extremely difficult to reproduce such
observables, even with the LDC. This is why we will here not be able
to constrain the prediction for the central exclusive production, but
on the contrary conclude that the uncertainties are larger than one
might expect.

The outline of this paper is as follows. First we recapitulate in
section \ref{sec:exclusive} the main points of the calculation of
Khoze, Martin and Ryskin. Then in section \ref{sec:LDC} we briefly
describe the Linked Dipole Chain model and explain how we use it to
obtain the central exclusive luminosity function. In section
\ref{sec:results} we present our results and compare them with the
calculation of Khoze, Martin and Ryskin, leading us to the conclusions
presented in section \ref{sec:conclusions}.

\section{Central exclusive production}
\label{sec:exclusive}

\FIGURE[t]{\epsfig{file=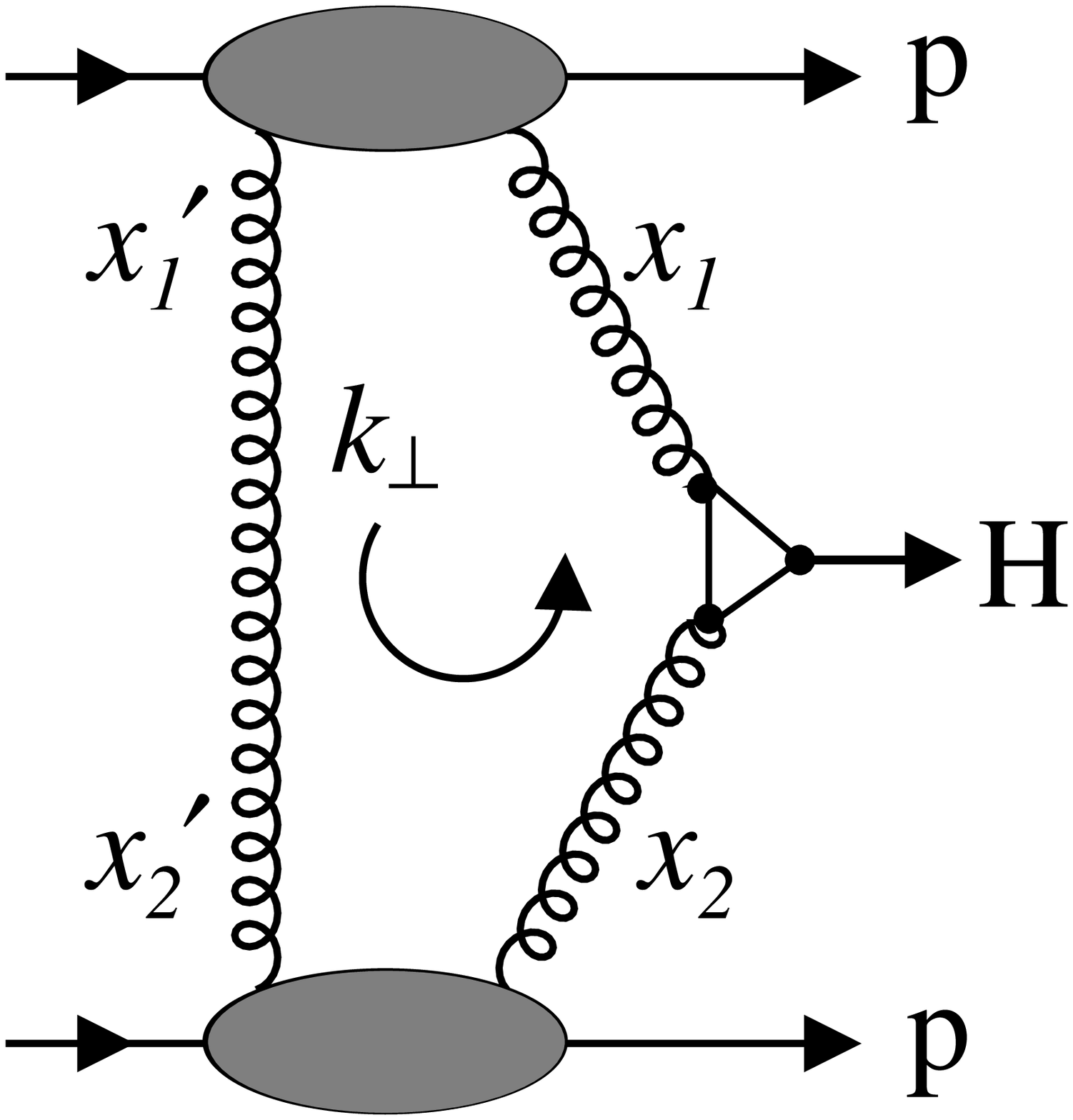,width=7cm}
  \caption{\label{fig:exdiagram} The basic diagram for exclusive production
    of the Higgs boson}}

The general idea for central exclusive production of a scalar particle
such as the Higgs boson\footnote{We will in the following talk only
  about the Higgs, but note that the formalism is valid for the
  production of any scalar particle.} is that two gluons with no net
quantum number fuse into a Higgs via the standard heavy quark triangle
diagram, whereas another semi-hard gluon guarantees that there is no
net colour flow between the protons. This is shown in figure
\ref{fig:exdiagram}, where it is also indicated that the exchanged
semi-hard gluon should also compensate the transverse momentum
$k_\perp$ of the gluons producing the Higgs, so that the protons are
scattered with little or no transverse momenta.

Several types of radiation can destroy the diffractive character of
the interaction.  An additional gluon or quark which destroys the
color singlet can be emitted by one of the gluons. For additional
gluons of $q_\perp>k_\perp$ this will be taken care of by a
\emph{hard} survival probability given by a Sudakov form factor (see
\eqref{eq:mishasudakov} below) which guarantees that no gluon or quark
with $q_\perp$ between $k_\perp$ and the hard scale given by $M$ is
emitted.

In principle there is also a probability of emitting a gluon of
transverse momentum squared less then $k_\perp$ and this probability
diverges for small $q_\perp$. However, the $k_\perp$ here acts as an
effective cut off since a gluon with a wavelength larger than
$1/k_\perp$ will not be able to resolve the individual colour flow of
the two gluons, but will only see a color singlet being exchanged.

Another process which reduces the number of diffractive events is
additional soft rescattering of the spectator partons. This is taken
care of by a soft survival probability, $S^2$, the value of which can
be be estimated by several different models. Here we will use the same
estimates as in \cite{Khoze:2001xm} where $S^2$ is taken to be 0.045
for the Tevatron and 0.02 for LHC.

Finally we must make sure that the protons remain intact, which gives
us a suppression depending on the momentum transfer to each the protons,
$t=(p_i-p_f)^2$:
\begin{eqnarray*}
    P = e^{b(p_i-p_f)^2}.
\end{eqnarray*}
This momentum transfer will be integrated over, giving a suppression
factor $1/b^2$, and we will here use the same value as in
\cite{Khoze:2001xm}: $b=4$~GeV$^{-2}$.

The exclusive cross section of $pp\rightarrow ppH$ can be factorized
into the form
\begin{eqnarray*}
  \sigma = \int \hat{\sigma}_{gg\rightarrow H}(M^2)
  \frac{\delta^2 {\cal L}}{\delta y \delta\ln{M^2}}
  dyd\ln{M^2}
\end{eqnarray*}
where $ \hat{\sigma} $ denotes the basic ${gg\rightarrow H}$
cross section and
\begin{eqnarray}
  \label{eq:kmrlum}
  L(M,y)&=&\frac{\delta^2 {\cal L}}{\delta y \delta \ln{M^2}}\\
  &=& S^2 \left[\frac{\pi}{(N_c^2-1)b}\int^{M^2/4} \frac{dk_\perp^2}{k_\perp^4}
    f_g(x_1,x_1',k_\perp^2,M^2/4)
    f_g(x_2,x_2',k_\perp^2,M^2/4)\right]^2\nonumber
\end{eqnarray}
with $x_{1,2}=e^{\pm y}M/E\sub{cm}$, is the luminosity function for
producing two gluons attached to the central process at rapidity $y$
and mass $M$ of the Higgs. In principle one should be using an
off-shell version of $\hat{\sigma}$ (see eg.\ \cite{Hautmann:2002tu})
which then would have a \kT\ dependence, hence breaking the
factorization, but we shall find below that the main contribution
comes from rather small \kT\ and at least for large masses the
factorization should hold.

The equation for the luminosity contains the off-diagonal unintegrated
gluon densities, $f(x,x',k_\perp^2,\mu^2)$. These should be
interpreted as the amplitude related to the probability of finding two
gluons in a proton carrying equal but opposite transverse momentum,
$k_\perp$, and carrying energy fractions $x$ and $x'$ each, one of
which is being probed by a hard scale $\mu^2$. To obtain these density
functions in \cite{Khoze:2001xm} the two step procedure presented in
\cite{Martin:2001ms} was used. First they obtain the odPDF from the
standard gluon PDF, in the here relevant limit of $x'\ll x$:
\begin{equation}
  \label{eq:pdf2odpdf}
  H(x,x',\mu^2)\approx R_g xg(x,\mu^2).
\end{equation}
The $R_g$ factor depends on the $x$-behavior of the PDF, so that for
$xg(x,\mu^2)\propto x^{-\lambda}$\cite{Shuvaev:1999ce},
\begin{equation}
  \label{eq:Rg}
  R_g=\frac{2^{2\lambda+3}}{\sqrt{\pi}}
  \frac{\Gamma(\lambda+5/2)}{\Gamma(\lambda+4)}
  \approx 1+0.82\lambda+0.56\lambda^2 +{\cal O}(\lambda^3)
\end{equation}
This factor can be taken approximately constant and we will then use
the values quoted in \cite{Khoze:2001xm}: $1.2$ for the LHC and $1.4$
at the Tevatron. We note, however, that it could also be taken to
depend on both $x$ and $\mu^2$ by using\footnote{Which was actually
  done in \cite{Khoze:2001xm} to obtain the luminosities
  \cite{ryskin:priv}.} $\lambda=d\ln{xg(x,\mu^2)}/d\ln(1/x)$.

In the next step it is assumed that the oduPDF can be obtained from
the odPDF in the same way as the uPDF can be obtained from the
standard PDF. In the latter case one can use the KMR prescription
introduced in \cite{Kimber:2001sc}, where
\begin{equation}
  \label{eq:pdf2updf}
  G(x,k_\perp^2,\mu^2)\approx
  \frac{d}{d\ln k_\perp^2}\left[xg(x,k_\perp^2) T(k_\perp^2,\mu^2)\right],
\end{equation}
which then corresponds to the probability of finding a gluon in the
proton with transverse momentum $k_\perp$ and energy fraction $x$ when
probed with a hard scale $\mu^2$. T is here the survival probability
of the gluon given by the Sudakov form factor,
\begin{equation}
  \label{eq:mishasudakov}
  -\ln T(k_\perp^2,\mu^2) =
  \int_{k_\perp^2}^{\mu^2}\frac{dq_\perp^2}{q_\perp^2}
  \frac{\alpha_S(q_\perp^2)}{2\pi}\int_{0}^{\frac{\mu}{\mu+k_\perp}}
  dz\left[zP_g(z) + n_f P_q(z)\right].
\end{equation}
To get the oduPDF one then starts from \eqref{eq:pdf2odpdf} and get by
analogy in the limit $x'\ll x$
\begin{equation}
  \label{eq:odpdf2odupdf}
  f_g(x,x',k_\perp^2,\mu^2)\approx
  \frac{d}{d\ln k_\perp^2}\left[R_g xg(x,k_\perp^2)
    \sqrt{T(k_\perp^2,\mu^2)}\right],
\end{equation}
where the square root of the Sudakov comes about because only one of
the two gluons are probed by the hard scale.

\FIGURE[t]{\epsfig{file=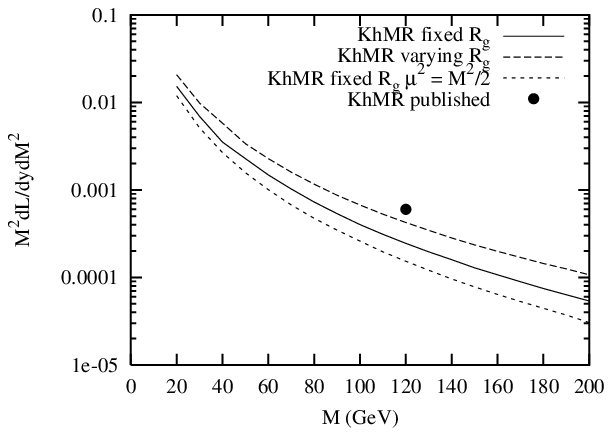,width=10cm}
  \caption{\label{fig:kmrlum} The exclusive luminosity as a
    function of $M$ for fixed rapidity, $y=0$, at the LHC, as
    calculated according to \eqsref{eq:kmrlum} and
    (\ref{eq:odpdf2odupdf}) with fixed $R_g=1.2$ (full line) and with
    varying $R_g$ according to \eqref{eq:Rg} (long-dashed line). The
    point is the the value quoted in \cite{Khoze:2001xm}. Short-dashed
    line is the same as the full line but using the scale
    $\mu^2=M^2/2$ rather than $\mu^2=M^2/4$ in the oduPDFs.}}

In figure \ref{fig:kmrlum} we show our calculation of the luminosity
function for central rapidity at the LHC using \eqref{eq:kmrlum}. We
use both a constant $R_g=1.2$ and a varying one according to
\eqref{eq:Rg} and we find that the treatment of $R_g$ does make a
difference. The latter alternative is closer to the result
\cite{Khoze:2001xm}, but it is not exactly the same due to differences
in the handling of the $\alpha_S$ in the Sudakov and the lower limit
in the integral of \eqref{eq:kmrlum}. We use a leading order
$\alpha_S$ and the cutoff is taken to be the input scale of the MRST99
(central-g L300-DIS) \cite{Martin:1999ww} used as the starting PDF in
\eqref{eq:pdf2odpdf}.

We note that the scale used in the oduPDFs in \cite{Khoze:2001xm} is
$\mu^2=M^2/4$ rather than the somewhat more natural one $\mu^2=M^2$.
Although we realize that in a leading order calculation like this the
scale choice is somewhat ambiguous. In figure \ref{fig:kmrlum} we show
that the scale choice in fact makes a rather big difference, the
luminosity function is reduced by up to 50\% by increasing the scale a
factor of two.

\section{The Linked Dipole Chain Model}
\label{sec:LDC}

We will here only describe the main characteristics of the LDC model
and instead refer the reader to refs.\ 
\cite{Andersson:1996ju,Andersson:1998bx,Gustafson:2002jy} for a more
detailed description. The Linked Dipole Chain model is a reformulation
and generalization of the CCFM evolution for the uPDFs. CCFM has the
property that it reproduces BFKL
evolution\cite{Kuraev:1977fs,Balitsky:1978ic} for asymptotically large
energies (small $x$) and is also similar to standard DGLAP evolution
\cite{Gribov:1972ri,Lipatov:1975qm,Altarelli:1977zs,Dokshitzer:1977sg}
for large virtualities and larger $x$. It does this by carefully
considering coherence effects between gluons emitted from the
evolution process, allowing only gluons ordered in angle to be emitted
in the initial state, and thus contribute to the uPDFs, while
non-ordered gluons are treated as final state radiation off the
initial state gluons.

\FIGURE[t]{
\begin{picture}(200,230)(-10,0)
\ArrowLine(10,15)(50,15)
\Text(20,25)[]{\large $proton$}
\Line(50,20)(100,20)
\Line(50,15)(100,15)
\Line(50,10)(100,10)
\Line(50,20)(80,40)
\GOval(50,15)(10,7)(0){1}
\Text(60,35)[]{\large $k_{0}$}
\Text(130,40)[]{\large $q_{1}$}
\Line(80,40)(120,40)
\Line(80,40)(95,70)
\Text(80,55)[]{\large $k_{1}$}
\Text(145,70)[]{\large $q_{2}$}
\DashLine(87.5,55)(110,55){2}
\Text(118,55)[]{\large $q'_{1}$}
\Text(90,85)[]{\large $k_{2}$}
\Text(155,100)[]{\large $q_{3}$}
\Line(95,70)(135,70)
\Line(95,70)(105,100)
\DashLine(100,70)(120,80){2}
\Line(105,100)(145,100)
\Line(105,100)(110,130)
\Line(110,130)(150,130)
\Line(110,130)(110,160)
\Line(110,160)(150,160)
\DashLine(120,160)(140,150){2}
\DashLine(130,155)(145,155){2}
\ArrowLine(20,220)(70,200)
\Text(166,160)[]{\large $q_{n+1}$}
\Text(100,145)[]{\large $k_{n}$}
\Text(20,210)[]{\large $lepton$}
\Text(100,185)[]{\large $q_{\gamma}$}
\ArrowLine(70,200)(120,220)
\Photon(70,200)(110,160){3}{5}
\end{picture}

  \caption{\label{fig:fanDIS} A fan diagram for a DIS event. 
    The quasi-real partons from the initial-state radiation are denoted
    $q_i$, and the virtual propagators $k_i$. The dashed lines denote
    final-state radiation.}}

The LDC model is based on the observation that the dominant features
of the parton chains are determined by a subset of emitted gluons,
which is ordered in both light-cone components, $q_+$ {\it and} $q_-$,
(which implies that they are also ordered in angle or rapidity $y$)
and with $q_{\perp i}$ satisfying the constraint
\begin{equation}
  q_{\perp i} > \min(k_{\perp i},k_{\perp,i-1}),
  \label{eq:ldccut}
\end{equation}
where $q$ and $k$ are the momenta of the emitted and propagating
gluons respectively as indicated in figure \ref{fig:fanDIS}.  In LDC
this subset (called ``primary'' gluons, or the backbone of the chain)
forms the chain of initial state radiation, and all other emissions
are treated as final state radiation.

This redefinition of the separation between initial- and final-state
implies that one single chain of initial-state emissions in the LDC
model corresponds to a whole set of CCFM chains.  As was shown in
ref.~\cite{Andersson:1996ju}, when summing over the contributions from
all chains of this set, the resulting equations for the primary chains
is greatly simplified. In particular the so-called non-eikonal form
factors present in the CCFM splitting functions do not appear
explicitly in LDC. The LDC formulation can also be easily made
forward-backward symmetric, so that in DIS, the evolution can be
equally well formulated from the virtual photon side or from the
proton side.

In the small-$x$ limit, keeping only the $1/z$ term of the gluon
splitting function we can write the perturbative part of the gluon
uPDF as the sum of all possible chains ending up with a gluon at a
certain $x$ and $\kT^2$
\begin{equation}
  \mathcal{G}(x,k_\perp^2) \sim \sum_n \int \prod^n \bar{\alpha} 
  \frac{dz_i}{z_i} \frac{d^2 q_{\perp i}}{\pi q_{\perp i}^2} 
  \theta(q_{+,i-1} -q_{+ i}) \theta(q_{- i} -q_{-,i-1}) 
  \delta(x-\Pi z_i) \delta(\ln k_\perp^2/k_{\perp n}^2),
  \label{eq:fq}
\end{equation}
where $\abar=3\alpha_s/\pi$. For finite $x$ it is straight forward to
add not only the $1/(1-z)$ to the gluon splitting function, as is also
done in CCFM, but also to include the full splitting function with
non-singular terms. The $z=1$ pole then needs to be regularized with a
Sudakov form factor $\Delta\sub{S}$ of the form
\begin{equation}
  \label{eq:yurisudakov}
  \ln\Delta\sub{S} = - \int\frac{dq_\perp^2}{q_\perp^2}
  \frac{\alpha_s}{2\pi}zdzP\sub{gg}(z)
  \Theta\sub{order},
\end{equation}
where $\Theta\sub{order}$ limits the integration to the phase space
region where initial-state emissions are allowed according to LDC. It
is also straight forward to add quarks in the evolution with the
appropriate modification of the Sudakov form factors.

The LDC model can easily be implemented in an event generator which is
then able to generate complete events in DIS with final state
radiation added according to the dipole cascade model
\cite{Gustafson:1986db,Gustafson:1988rq} and hadronization according
to the Lund model \cite{Andersson:1983ia}. In addition, the
perturbative form of the uPDF in \eqref{eq:fq} needs to be convoluted
with non-perturbative input PDFs, the form of which are fitted to
reproduce the experimental data on $F_2$. This has all been
implemented in the \ldcmc program \cite{Kharraziha:1998dn}, and the
resulting events can be compared directly to experimental data from
eg.\ HERA.  One of the most important observables is the rate of
forward jets which is sensitive to parton evolution with unordered
tranverse momenta, which is modeled by BFKL, CCFM and LDC, but is not
allowed DGLAP. This observable should also be especially sensitive to
the actual \kT distribution of gluons in the proton. It turns out that
the forward jet rates can indeed be reproduced by \ldcmc (as well as
with the CCFM event generator \cascade \cite{Jung:2001hx}) but only if
only gluons are included in the evolution and if non-singular terms
are excluded from the gluon splitting function \cite{Anderson:2002cf}.
So far there is no satisfactory explanation for this behavior.

\FIGURE[t]{
  \epsfig{file=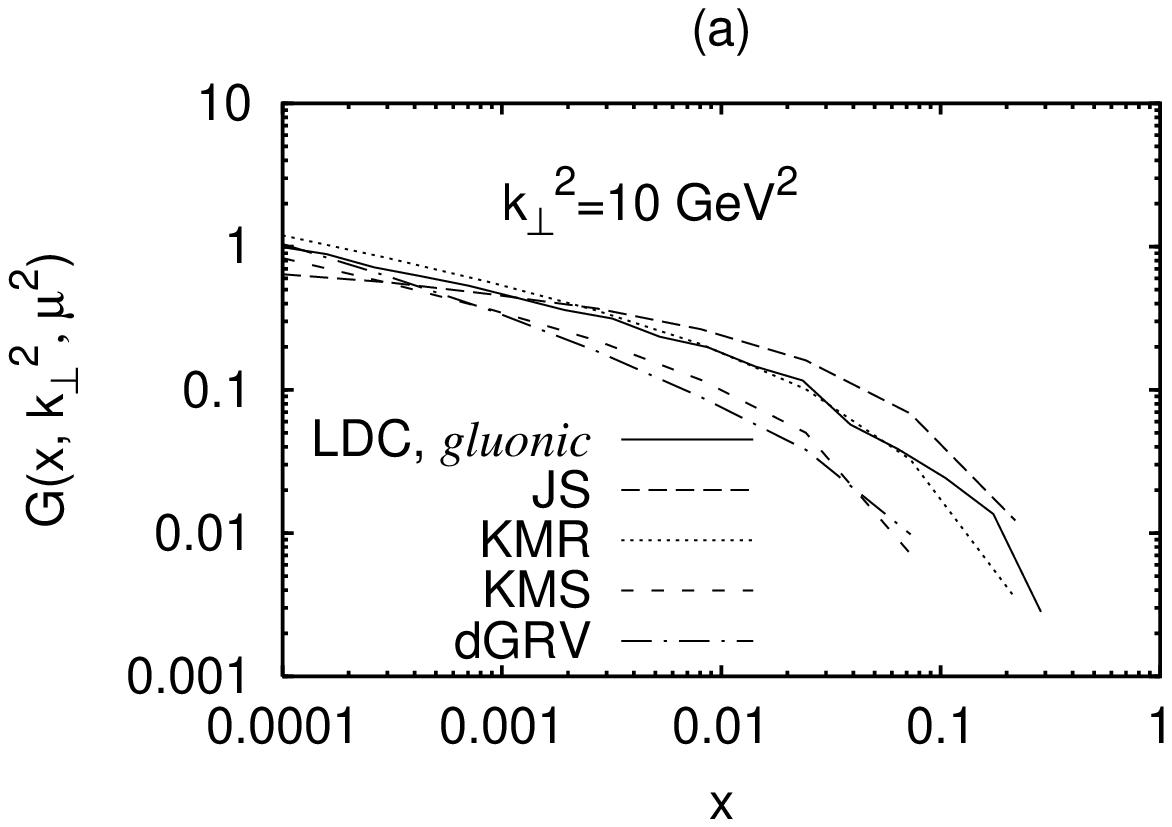,width=7.45cm}\epsfig{
    file=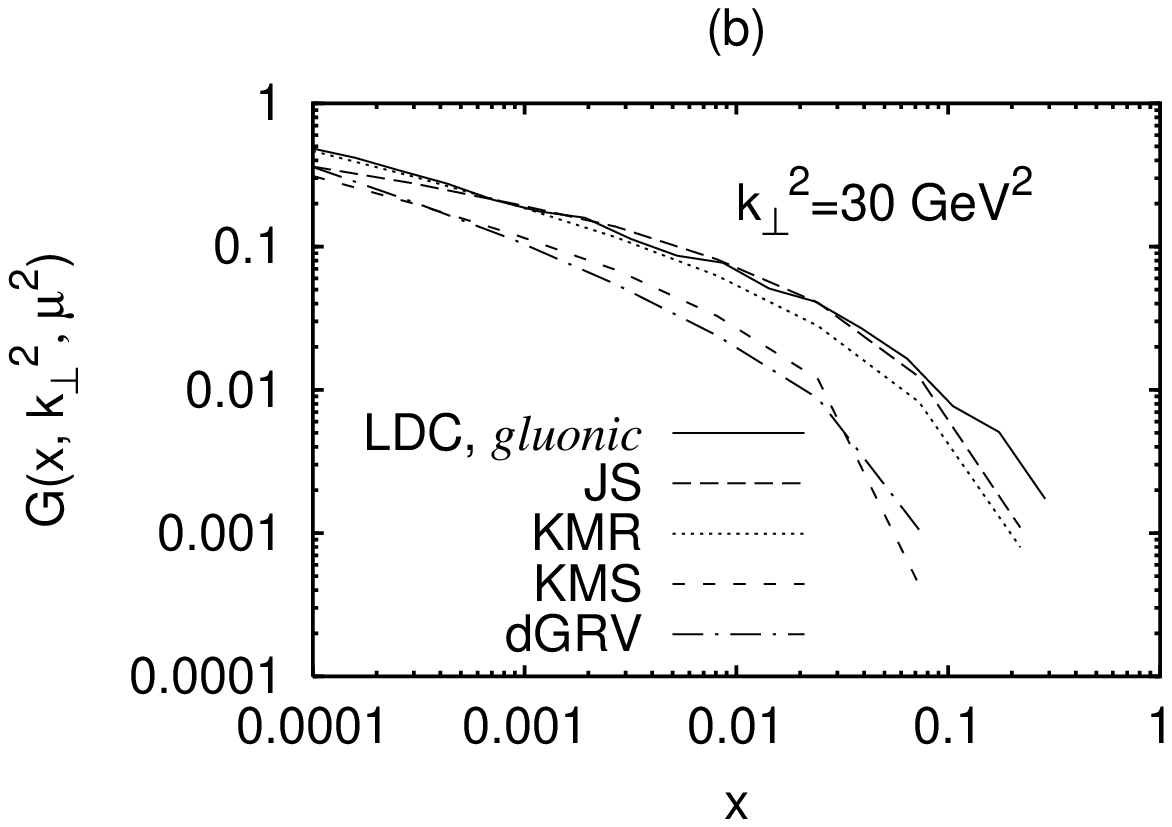,width=7.45cm}

  \vspace*{-5mm}

  \caption{\label{fig:unint-comp-1sc} The LDC \textit{gluonic}
    unintegrated gluon distribution function (full curve), compared to
    the corresponding results of JS \cite{Jung:2000hk} (long-dashed
    curve), KMR \cite{Kimber:2001sc} (dotted curve), KMS
    \cite{Kwiecinski:1997ee} (short-dashed curve) and a simple
    derivative of the GRV \cite{Gluck:1998xa} PDF parameterization
    (dash-dotted curve) as functions of $x$ for (a) $\kT^2=
    10$~GeV$^2$ and (b) $\kT^2= 30$~GeV$^2$.  Results for the
    two-scaled functions, LDC, JS and KMR, with $\mu=2 \kT$, are shown
    together with the 1-scaled distribution functions of KMS and
    dGRV.}}

The LDC gluon uPDF has been extracted by generating a DIS events with
\ldcmc and measuring the gluon density as described in
\cite{Gustafson:2002jy}. The density depends on two scales, \kT and a
scale, $\bar{q}$, related to the maximum angle allowed for the emitted
gluons, which is related to the virtuality $\mu^2$ of the hard
sub-process. In LDC, similarly to the KMR prescription, the uPDF
approximately factorizes into a single scale uPDF and a Sudakov form
factor:
\begin{equation}
  \label{eq:fact}
  G(x,\kT^2,\mu^2)\approx G(x,\kT^2)\times\Delta_S(\kT^2,\mu^2).
\end{equation}

This density can then be compared to other approaches and one finds
that the results are quite varying as the examples in figure
\ref{fig:unint-comp-1sc} shows. Even looking only at the proper
two-scale uPDFs, factors of two difference are not uncommon.

Due to the \kT-unordered nature of the LDC evolution, the relationship
between the uPDF and the standard gluon density is different from
\eqref{eq:pdf2updf}, as the integrated gluon at a scale $\mu^2$ also
receives a contribution, although suppressed, from gluons with
$k_\perp>\mu$, and in \cite{Gustafson:2002jy} the following expression
was obtained:
\begin{eqnarray}
  \label{eq:updf2pdf}
  xg(x,\mu^2)&=&G(x,k_{\perp0}^2)\Delta_S(k_{\perp0}^2,\mu^2)\\
  &+&\int_{k_{\perp0}^2}^{\mu^2}\frac{dk_\perp^2}{k_\perp^2}
  G(x,k_\perp^2)\Delta_S(k_\perp^2,\mu^2) +
  \int_{\mu^2}^{\mu^2/x}\frac{dk_\perp^2}{k_\perp^2}
  G(x\frac{k_\perp^2}{\mu^2},k_\perp^2)\frac{\mu^2}{k_\perp^2}\nonumber
\end{eqnarray}

To obtain the off-diagonal densities needed for the exclusive
luminosity function, we assume that a similar approximation can be
made as for the KMR densities, that is, in the limit of very small $x'$
\begin{equation}
  \label{eq:oduLDC}
  f_g\sup{LDC}(x, x', \kT^2, \mu^2)\approx
  R_g(x,\kT^2)G(x,\kT^2)\sqrt{\Delta_S(\kT^2,\mu^2)}.
\end{equation}
The square root of the Sudakov form factor is used, since only one of
the gluons couples to the produced Higgs at the high scale. We will
use both a fixed $R_g$ factor as in section \ref{sec:exclusive} and
the one which depends explicitly on the $x$-dependence of the diagonal
PDF taken at the relevant scale. It is currently not quite clear to us
how large the uncertainties are in this procedure and we come back to
it in a future publication.

The LDC uPDFs are only defined down to a cutoff, $k_{\perp0}$, below
which we will use the non-perturbative input density, $g_0$, and arrive
at the following expression for the exclusive luminosity function:
\begin{eqnarray}
  \label{eq:ldclum}
  L=S^2\left[
    \frac{\pi}{(N_c^2-1)b}\rule{0mm}{7mm}\right.&&
  \!\!\!\!\!\left(\rule{0mm}{7mm}
    \frac{1}{k_{\perp0}^2}R_g(x_1,k_{\perp0}^2)
    g_0(x_1,k_{\perp0}^2)\Delta_S(k_{\perp0}^2,M^2)
    R_g(x_2,k_{\perp0}^2)g_0(x_2,k_{\perp0}^2)
     +\right.\nonumber\\
  & &\!\!\!\!\!\!\!\!\!\!\!\!\!\!\left.\left.
      \int_{k_{\perp0}^2}^{M^2}\frac{dk_{\perp}^2}{k_{\perp}^4}
      R_g(x_1,k_{\perp}^2)
      G(x_1,k_{\perp}^2)\Delta_S(k_{\perp}^2,M^2)
      R_g(x_2,k_{\perp}^2)G(x_2,k_{\perp}^2)
      \right)\right]^2.
\end{eqnarray}

Comparing with \eqref{eq:kmrlum} we note that, besides the
different form of the oduPDFs, the scale and the integration limit is
taken to be $M^2$ rather than $M^2/4$. The exact value of the
integration limit is not very important, but the scale in the Sudakov
form factor is. In fact, the form of the Sudakov form factor is also
different. We use
\begin{equation}
  \label{eq:sudLDC}
  \ln\Delta_S(k_\perp^2,M^2)=
  -\int_{k_\perp^2}^{M^2}\frac{dk_\perp^2}{k_\perp^2}
  \frac{\alpha_s}{2\pi}
  \int_0^{1-k_\perp/M}dz\left[zP_g(z) + \sum_qP_q(z)\right],
\end{equation}
which corresponds to the actual no-emission probability in the phase
space region up to the rapidity of the produced Higgs from the
incoming gluon. The different integration region in
\eqref{eq:mishasudakov} as well as the different scale used means that
the Sudakov suppression in that case is weaker as shown in figure
\ref{fig:sudKMR}. The difference is not very large, but since the
factor comes in squared in the luminosity function for $\kT$ of a
couple of GeV the difference can easily become larger than a factor
two.

\FIGURE[t]{ \epsfig{file=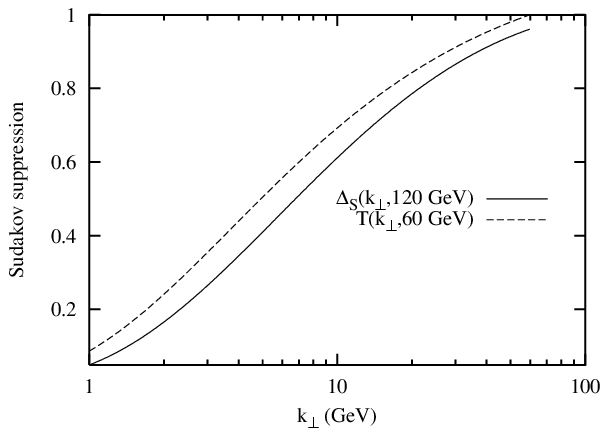,width=10cm}
  \caption{\label{fig:sudKMR} The Sudakov form factor used in the LDC
    calculation (full line, \eqref{eq:sudLDC}) compared to the one
    used by KhMR (dashed line, \eqref{eq:mishasudakov})}}

One of the main differences between the LDC uPDFs and the KMR ones is
that the evolution of former includes emissions with transverse
momenta which may be larger than the \kT\ for the probed gluon. This
is, of course, kinematically allowed but should be rather suppressed.
In any case, it is not clear how to handle such emissions when
calculating the off-diagonal densities in \eqref{eq:oduLDC}. Below
we shall therefore also use an alternative LDC uPDF where the
transverse momentum in the evolution has been limited to be below \kT.

\section{Results}
\label{sec:results}

In the following we shall present calculations for the exclusive
luminosity using three different parameterizations of the LDC uPDF.
The three options differs in the way they treat non-leading effects in
the evolution and will be referred to as \standard,
\gluonic and \leading as described in
\cite{Gustafson:2002jy}:
\begin{itemize}
\item \standard is obtained with the full LDC evolution including the
  full splitting functions for both gluons and quarks.  This option
  does not describe forward jets very well, but it gives an excellent
  description of $F_2$ data.
\item \gluonic is obtained by using only gluons in the LDC
  evolution, but with the full splitting function. This option does
  not describe $F_2$ data as well, especially not at large $x$, but it
  agrees better with standard parameterizations of the integrated
  gluon PDF.
\item \leading is obtained by using only gluons in the LDC
  evolution and only the singular terms of the gluon splitting
  function. Among the three it is the one which describes inclusive
  data the worst, on the other hand it is the only one which is able
  to describe the large rate of forward jets measured at HERA.
\end{itemize}
Clearly, none of these options are in perfect agreement with data, but
we will use them here as a parameterization of our ignorance when it
comes to unintegrated gluon densities.

\FIGURE[t]{\epsfig{file=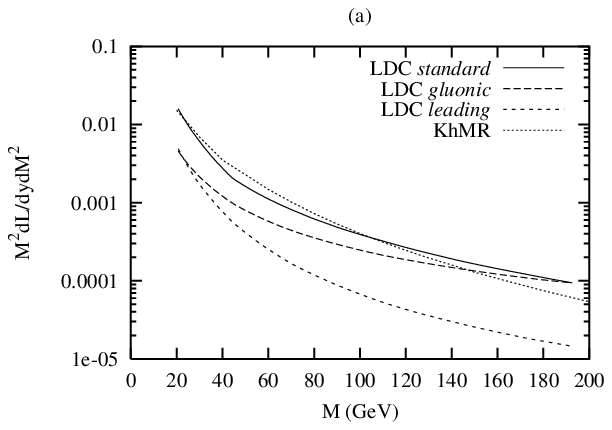,width=7.45cm}\epsfig{
    file=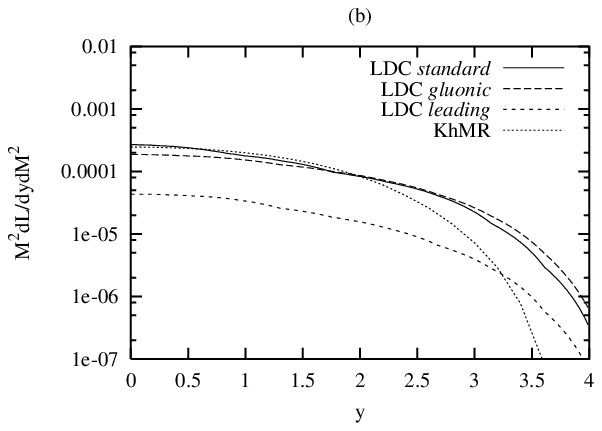,width=7.45cm}
  \caption{\label{fig:ldclum} The exclusive luminosity as a
    function of $M$ for fixed rapidity, $y=0$ (a) and as a function of
    rapidity for fixed mass $M=120$~GeV, at the LHC, as calculated
    according to \eqref{eq:ldclum}. Full line is \textit{standard},
    long-dashed line is \textit{gluonic} and short dashed line is
    \textit{leading}. As comparison the calculation based on KhMR is
    shown with a dotted line.}}

\FIGURE[t]{\epsfig{file=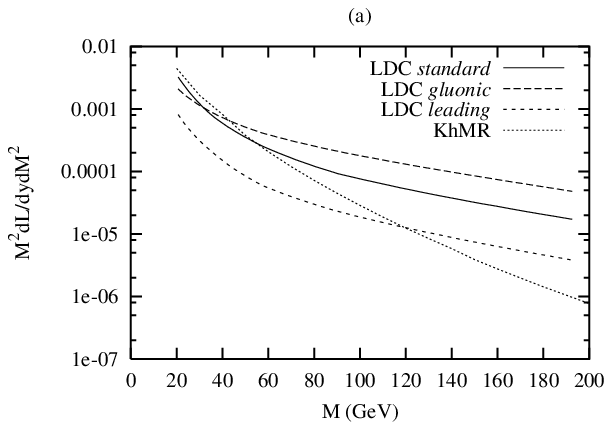,width=7.45cm}\epsfig{
    file=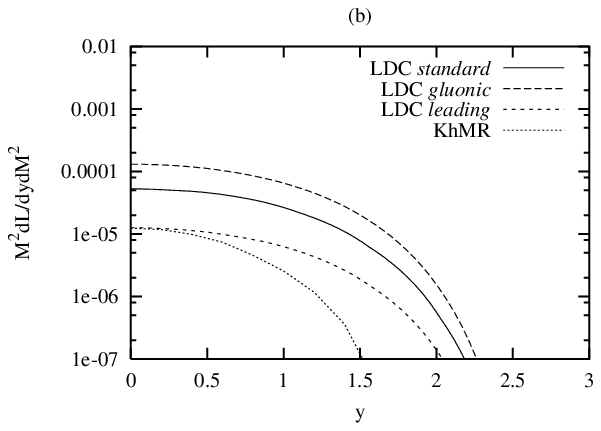,width=7.45cm}
  \caption{\label{fig:tevlum} The exclusive luminosity as a
    function of $M$ for fixed rapidity, $y=0$ (a) and as a function of
    rapidity for fixed mass $M=120$~GeV, at the Tevatron, as
    calculated according to \eqref{eq:ldclum}. The lines are the same
    as in figure \ref{fig:ldclum}.}}

In figure \ref{fig:ldclum} and \ref{fig:tevlum} we present our
calculations of the luminosity function for the LHC and Tevatron
respectively, using \eqref{eq:ldclum} (with fixed $R_g=1.2$ and $1.4$
respectively). We find that the three options for the LDC evolution
give very different results. At the LHC the \textit{standard} is
fairly close to the results obtained with the KhMR calculation, while
the result for \textit{leading} is up to a factor ten below. We note
that for large rapidities in figure \ref{fig:ldclum}b the differences
between LDC and KhMR is larger also in shape, but this is close to the
phase space limit, where one of the gluons carry a large fraction of
the proton momentum and in this region the LDC parameterizations are
less well constrained. The same effect is visible at the Tevatron in
figure \ref{fig:tevlum} where again the energy fractions are larger,
especially for high masses.

\FIGURE[t]{\epsfig{file=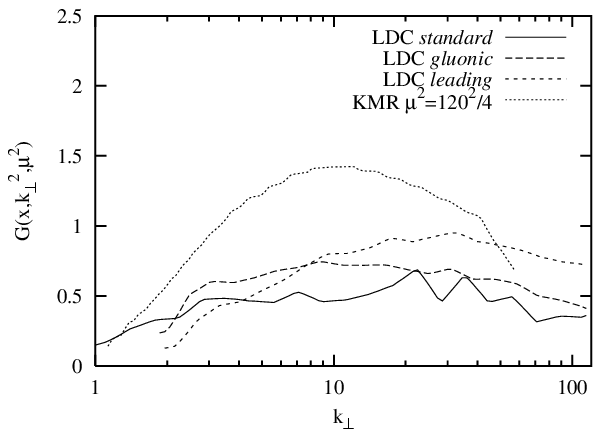,width=10cm}
  \caption{\label{fig:updf} The LDC unintegrated gluon density as a
    function of \kT\ for $\mu^2=(120\mbox~GeV)^2$ and $x=120/14000$
    (relevant for the luminosity function in \eqref{eq:ldclum} for
    $M=120$~GeV and $y=0$).  For comparison the KMR uPDF is shown for
    the same $x$ but for $\mu^2=(120/2\mbox~GeV)^2$ (relevant for
    \eqref{eq:kmrlum}). The lines are the same as in figure
    \ref{fig:ldclum}..}}

The large difference between the three LDC options may seem
surprising, especially since the standard integrated gluon PDF is
generally higher for \textit{leading} than for the other two.
The explanation can be found by studying the $k_\perp$-dependence of
the uPDF presented in figure \ref{fig:updf}. Here we see that \leading
has a harder \kT\ spectrum than the the other two options and that all
LDC densities have a flatter spectrum than the KMR one (which is shown
at a lower scale corresponding to the one used in the luminosity
function). This should be expected since the \leading also produces
more forward jets (in agreement with what is observed experimentally)
and hence should give larger \kT-fluctuations. It turns out that the
luminosity function is mostly sensitive to the uPDFs at \kT-values of
around $2-3$~GeV, since smaller and larger values are suppressed by
the Sudakov form factor and the $1/\kT^4$ factor respectively. Even if
the differences in this region is not very large, the uPDF enters to
the power four in the luminosity function, thus enhancing the
differences (the difference between LDC and KMR is diminished since
the square root of the Sudakov formfactor affect the LDC more than the
KMR).

\FIGURE[t]{\epsfig{file=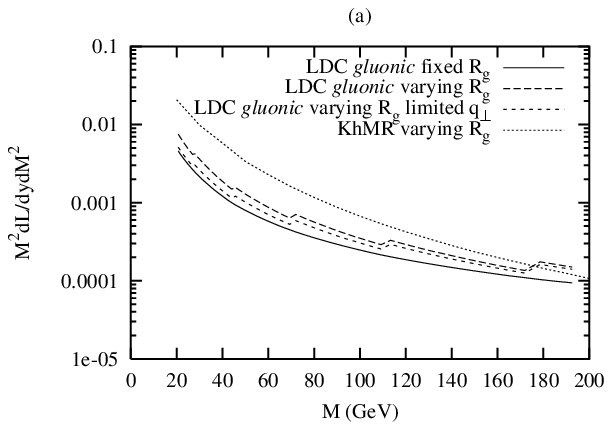,width=10cm}
  \caption{\label{fig:ldclumr} The exclusive luminosity as a
    function of $M$ for fixed rapidity, $y=0$ at the LHC, as
    calculated according to \eqref{eq:ldclum} using the \gluonic uPDF
    of LDC. Full line is with a fixed $R_g=1.2$, long-dashed and
    short-dashed lines are with varying $R_g$ but the latter uses a
    modified version of \gluonic where the transverse momentum in the
    evolution has been limited. As comparison the calculation based on
    KhMR with a varying $R_g$ is shown with a dotted line.}}

To investigate the uncertainties involved in going from the LDC uPDFs
to the oduPDFs in \eqref{eq:oduLDC} we show in figure
\ref{fig:ldclumr} the difference between using a fixed $R_g$ factor
and a varying one according to \eqref{eq:Rg}\footnote{The line is here
  a bit jagged due to the limited statistics in the Monte Carlo
  extraction of the LDC uPDF.} with
$\lambda=d\ln{G(x,\kT^2)}/d\ln(1/x)$. Comparing with figure
\ref{fig:kmrlum}, we find that the differences are of the same order.
We also show the effects of using an alternative version of the
\gluonic density where the transverse momentum in the evolution has
been limited to be below \kT. This will not only reduce the uPDF
somewhat, but it will also slow down the $x$-evolution, giving a
smaller $\lambda$ and hence a smaller $R_g$. As expected this effect
is quite small, but still noticeable especially at small masses (small
$x$).

\section{Conclusions}
\label{sec:conclusions}

The partonic evolution at small $x$ is one of the least understood
aspects of QCD. We know that in the limit of asymptotically large
energies, BFKL evolution and \kT-factorization should give the correct
description, but it is known that for finite energies there are large
sub-leading corrections, which are not yet fully under control.
Although inclusive Higgs production is not a small-$x$ process and
therefore well understood in terms of collinear factorization with
well constrained integrated gluon distributions, the exclusive
production considered here relies on the exchange of a small-$x$ gluon
and is very sensitive to the \kT\ distribution in the less constrained
unintegrated gluon distributions.

In this report we have described how we calculate the exclusive
luminosity function using the unintegrated gluon distributions
obtained within the LDC model, and we have found that different
options give widely different results. In particular we note that the
option which gives the best description of forward jet production at
HERA, which should be sensitive to the actual \kT-dependence of the
gluon in the proton, gives a result which is a factor ten smaller than
what was reported by Khoze, Martin and Ryskin in \cite{Khoze:2001xm}.
This option is in theory a worse approximation than the other two and
is similar to the double-log approximation discussed by the same
authors in \cite{Khoze:2000cy}, which was also shown to give a much
smaller result. Contrary to them, however, we do not dismiss the
\leading approximation, as experiments indicate that it better
describes the actual \kT distrubution of the gluon.

There are several uncertainties in our calculations. The relation
between the unintegrated gluon and the corresponding off-diagonal
unintegrated gluon density not formally derived, but just assumed to
be valid by analogy. The results are sensitive to the treatment of the
$R_g$ factor and the treatment of the \kT-unordered nature of LDC
evolution. The different options used for the LDC unintegrated
densities are in good agreement with different kinds of experimental
observables, but none of them agrees with all important observables. It
should also be noted that these densities were obtained through a fit
to $F_2$ data only, which is mainly concentrated at small scales. At
large scales which are important for reasonable values of the Higgs
mass ($\gtaet 120$~GeV) the densities are less constrained.

The conclusion of this paper is therefore not that the previous
calculations by Khoze, Martin and Ryskin is wrong in any way, but
rather that they may have underestimated the uncertainties due to the
unintegrated gluon density. We will not go so far as to say that the
uncertainties are as large as a factor ten, but we believe that they
are much larger than a factor of two. This does not mean that the
prospects of using tagged forward protons to try to find the Higgs or
other scalar particles at the LHC becomes less interesting, but our
current understanding of the small-$x$ sector of QCD clearly needs to
be improved before we can give reliable predictions for such
processes.

\section*{Acknowledgments}

We would like to thank Valery Khoze and Mikhail Ryskin for important
comments and discussions. In addition we have benefitted from
discussions with Brian Cox, Jeff Forshaw, Gösta Gustafson and James
Monk.

\bibliographystyle{utcaps}  
\bibliography{references,refs} 

\end{document}